# Echo Mappping of the BLR – A Critical Appraisal

Dan Maoz

*School of Physics & Astronomy and Wise Observatory, Tel-Aviv University, Tel-Aviv 69978, ISRAEL*

**Abstract.** As variability data have improved, evidence has accumulated that some of the basic assumptions made when attempting to carry out echo mapping of AGNs are invalid. The results of the major monitoring campaigns confirm beautifully the "big picture" of the echo paradigm, but the details of the emission line light curves cannot be accurately reproduced with only the simplest assumptions. I discuss possible solutions. I present some preliminary optical light curves from Wise Observatory for NGC 4151 during the December 1993 multi-satellite campaign. The optical data show a continuity with the phenomenon observed in the IUE data: There is less flickering with increasing wavelength, as well as a monotonically increasing variable component that dominates *more* with increasing wavelength. This complex continuum behavior may be the explanation for the peculiarities mentioned above, i.e., the ionizing continuum behaves differently from the observed continuum. I review some recent results on quasar emission line variability from the Steward-Wise PG quasar monitoring program, demonstrating that quasar emission lines respond to continuum variations on timescales of months. Preliminary analysis of the emission line lag in these quasars allows one to extend the observed AGN Radius–Luminosity relation to higher luminosities than previously feasible. Agreement with the expected $R \propto L^{1/2}$ relation is suggested, but the small masses implied (and hence high Eddington ratios) may pose a problem for thin accretion disk models. Finally, I criticize the trend to attribute significance to the details of transfer functions recovered by inversion techniques. I show, as an example, that the model emission-line light curves produced by convolving the 5-year continuum light curve of NGC 5548 with transfer functions peaked at zero or non-zero lag have differences much smaller that the uncertainties in the H$\beta$ light curve. Transfer functions of both kinds can reproduce the data equally well. I emphasize the need to use modeling, rather than inversion methods, in order to delineate the regions of parameter space allowed and ruled out by the data.

# 1. Echo Mapping Works, But There Are Problems

The past years have seen the execution of increasingly ambitious echo-mapping campaigns (see review by Peterson, this volume). As the data quality have improved, we have seen the basic prediction of the AGN photoionization model beautifully confirmed, namely, that emission-line light curves mimic the continuum behavior, but with a lag due to light-travel time effects. Probably most striking were (and still are) the IUE results for NGC 5548 in 1989 (Clavel et al. 1991). The gross structure of the light curves for the strong lines is just that expected.

Before long, however, various peculiarities were noticed in the NGC 5548 data, as well as in other objects: A time-variable lag (Netzer & Maoz 1990; Peterson et al. 1994), suggesting an evolving BLR, but with continuum amplitude or variation timescale possibly being the true independent variables; indications of non-linear or negative line response (Maoz 1992; Sparke 1993; Wanders & Horne 1994); different lags for different lines (Clavel et al. 1991), and different derived transfer function shapes for different lines (Krolik et al. 1991; Horne, Welsh & Peterson 1992; see, however, §4) interpreted as evidence for a stratified BLR and/or a complex BLR geometry.

The nonlinear behavior is most clearly visible in the NGC 5548 1989 light curve of CIV $\lambda1549$. The total energy in the third continuum "event" is much less than that of the previous two events and, if the line light curve were a linear convolution of the continuum light curve, would produce a correspondingly weak feature in the emission line light curve. Instead, the CIV flux rises to the same amplitude it had in response to the previous two events. This observed nonlinearity is indicating that one or more of our assumptions are not completely valid. A number of explanations are possible. The local response of a specific line need not be completely linear, i.e., the change in line flux from a BLR cloud may not be linearly proportional to the change in the ionizing flux, due to, e.g. temperature and optical depth effects. For example, if the response of a BLR cloud in the CIV line starts to saturate, high and low continuum fluxes could produce similar line fluxes. A related effect is the depletion of a given ion species when the ionization structure of a cloud changes following continuum changes, which can also manifest itself as negative line response. Goad, O'Brien & Gondhalekar (1993; see also contributions by Goad and by O'Brien in this volume) have shown that nonlinearities of this sort are, in fact, predicted even by simple photoionization models. It is unclear yet whether this effect is strong enough to reproduce the third event in NGC 5548.

One hint that the details of the response of a particular line are not the entire story behind the problem comes from examination of the *total* line flux of NGC 5548. As long as each BLR cloud remains optically thick to the ionizing radiation, the energy from each ionizing photon must come out as one line photon or another (or several). Therefore, the nonlinear aspects should disappear in the total emission line light curve.(This is not a new idea; see Blandford and McKee [1982], §II.a.) The 1989 NGC 5548 campaign data give us the opportunity to carry out this test, since all the major emission lines, from Ly$\alpha$ through H$\alpha$ were observed simultaneously for the entire 8-month period. Figure 1 shows the NGC 5548 total line light curve (work with B. Peterson and H. Netzer; see also Netzer 1993) superposed on the continuum light curve. The total line flux icludes the FeII + Balmer continuum complex (the "small blue bump") which constitutes about 1/3 of the total line flux (Maoz et al. 1993).

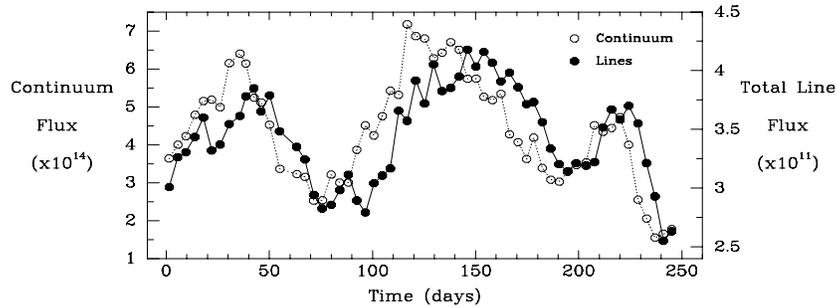

Figure 1. UV (1350Å) continuum light curve (empty circles, left vertical scale, in erg s$^{-1}$ cm$^{-2}$ Å$^{-1}$) for NGC 5548, and total observed emission-line flux (filled circles, right vertical scale, in erg s$^{-1}$ cm$^{-2}$ ), during the 1989 IUE monitoring campaign.

The third-event problem is considerably reduced in the total line light curve (Fig. 1), because the amplitude of the third-event response is much milder in the main components of the total line flux, Ly$\alpha$, the small blue bump, and H$\alpha$, than in CIV. The problem has, however, not disappeared. Figure 2 (top panel) shows the same total line curve (now shown as a jagged dashed line). Superposed on it is a reconstructed light curve (smooth solid line) obtained with a maximum entropy inversion that was forced to produce a transfer function that is monotonically decreasing (bottom panel). Maximum entropy methods applied to the NGC 5548 data have generally produced separate aliasing peaks in the transfer function which "conspire" with previous continuum events in order to ameliorate the third-event problem. Figure 2 shows that if such aliasing is not allowed, the amplitude of the third event is still much too high, even in the total line flux. The detailed physics of an individual emission line are therefore not the entire solution to the problem.

The remaining possibilities are:
1. The assumption that the BLR gas remains optically thick has broken down, and varying fractions of the BLR gas become optically thin as the continuum flux changes (see contributions by Shields and by Taylor);
2. The continuum behavior we see is not the ionizing continuum behavior that the BLR gas sees. This can come about if the continuum emission is not emitted isotropically, or if the ionizing continuum is not strictly linearly proportional to the observed continuum longward of the Lyman edge. Furthermore, the same observed continuum level at different epochs may not necessarily imply the same ionizing continuum level.

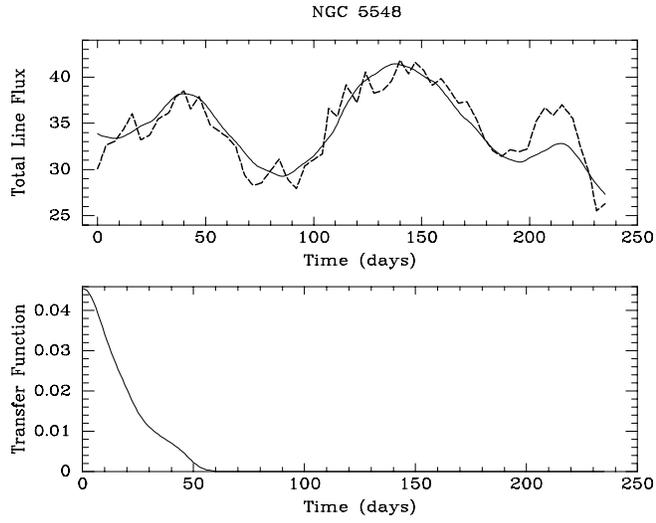

Figure 2. Top panel: Jagged dashed line is the total emission line light curve of NGC 5548, same as in Fig. 1. Bottom panel: Transfer function obtained by maximum entropy inversion of the total light curve with the UV continuum light curve, excluding the third "event". The smooth solid line in the top panel is the reconstructed emission line light curve obtained by convolving the continuum light curve with the transfer function in the bottom panel. The amplitude of the third event is poorly reproduced for any such monotonically decreasing transfer function, even though the total line flux ought to respond linearly to the continuum.

There is already evidence for possibility 2 in NGC 5548, in the well-established hardening of the continuum as it rises (e.g. Maoz et al. 1993). In the next section I will present some new evidence for such continuum "misbehavior" in another AGN, NGC 4151. If an imperfect correlation between the observed and ionizing continua is the source of the third-event problem (and the other problems mentioned above, but not discussed in depth), it could a pose a difficult problem for progress in echo mapping, as the ionizing continuum itself cannot be directly observed.

## 2. NGC 4151: Preliminary Optical Results

Perhaps the most impressive AGN monitoring campaign was successfully carried out in December 1993. The nearby Seyfert galaxy NGC 4151 was observed simultaneously by IUE, ASCA, ROSAT, and GRO for about 2 weeks. The IUE observations were continuous for 10 days (December 1-10), with approximately a one-hour sampling interval, giving us an unprecedented look at the high frequency behavior of a Seyfert nucleus (Crenshaw et al. 1994; see also contribution by Edelson in this volume). A number of ground-based observers carried out optical spectrophotometry, extending the wavelength coverage to the near IR. I present here some first-look results from the optical monitoring at Wise Observatory. Full results, including intercalibrated data from the other ground-based observations will appear in Kaspi et al. (1994).

ROSAT constraints dictated that NGC 4151 be observed in December, when it is a difficult ground-based target observable only in the last few hours of the night. At Wise Observatory we decided to make a concentrated effort for this unique multiwavelength opportunity. We began obtaining nightly spectra two weeks ahead of the satellites, and continued well into January 1994. The weather was cooperative, with no gaps in the sampling during the intensive 10-day period, and few gaps elsewhere. Spectra were obtained at Wise on 49 nights. We used our by-now traditional method of simultaneously observing, through a long and properly rotated slit, a field star with the AGN (e.g. Maoz et al. 1990, 1994). The star then serves as an excellent flux calibration standard, especially important in this object whose narrow-line region is extended, complicating the use of [OIII]$\lambda$5007 for calibration. The target was observed twice every night with a short wavelength setting, including H$\beta$, and twice with a long wavelength setting, including H$\alpha$. Each setting required renewed placing of the object in a different slit, and so constitutes a completely independent measurement. There is a large overlap in wavelength coverage between the two settings, and preliminary comparison of the results for a given night shows excellent repeatability.

Figure 3 shows the continuum lightcurves we measure in two different bands in our spectra, and compares them to the IUE light curves at two different bands. The light curves have been scaled and shifted so as to fit on the same plot. Two features are apparent: First, the trend observed in the IUE light curves, where the high-frequency power of the variations decreases with increasing wavelength, continues into the optical. Note how the 2800Å light curve seems like a smoothed

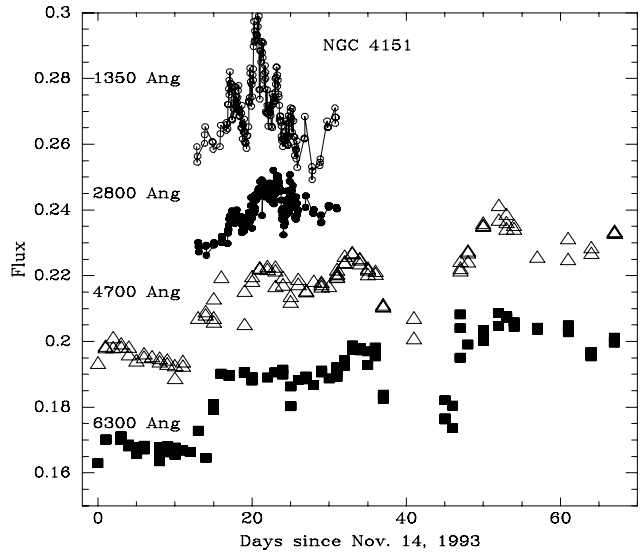

Figure 3. UV and optical continuum light curves for NGC 4151. The top two light curves are from the IUE SWP and LWP cameras, the bottom two are from blue and red spectra taken at Wise Observatory. Flux is in relative units and the individual light curves have been scaled and shifted vertically to fit on the same plot. Note how, as one goes to longer wavelengths, the variations become smoother and there is increasing dominance of a slow risisng component

version of the 1350Å light curve, there is still a hint of the same variation patterns in the 4700Å light curve, but the 6300Å light curve during the period of common coverage is almost completely smooth. Second, there is a long timescale monotonically increasing trend with several large jumps. The IUE coverage period is too short to see this trend clearly, but there is a hint that this slowly varying component becomes *more* dominant as one goes to longer wavelength.

Support for this suggestion of two variable components, a flickering UV one, with a fairly constant mean, and a slowly rising red one, comes from our measurements of H$\alpha$ and H$\beta$. Figure 4 shows the H$\alpha$ light curve. The light curve shows a slow monotonic rise, but this rise of 7% is much smaller than the corresponding rise in the total optical continuum level during this period (20-25%). Since we were using a wide (10″) slit that includes a large amount of stellar light from the galaxy, the true optical continuum variations were even larger. The non-response of the Balmer lines can be explained if the rising trend indeed disappears as one goes to the ionizing part of the continuum, and only the rapid flickering remains. The Balmer lines come from too large a region ($9 \pm 2$ lt-days; Maoz et al. 1991) to respond coherently to these fast variations.

Perhaps even more worrisome, from echo-mappers' point of view, than the clearly poor correlation between continuum behavior at different wavelengths is its *non-repetitive* behavior. A similar amplitude of optical continuum variations

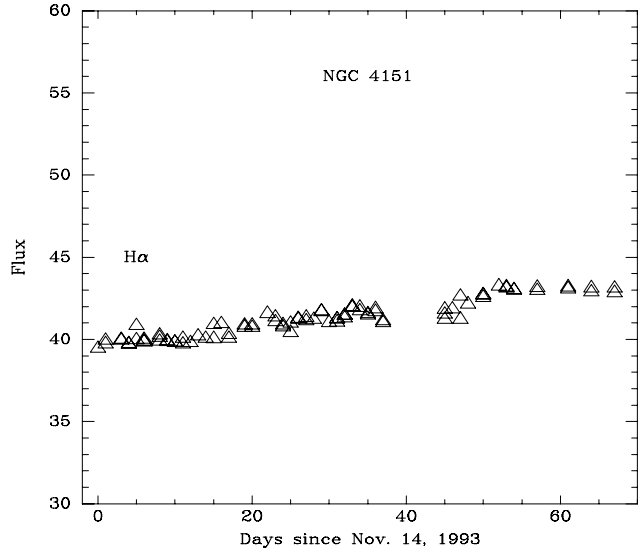

Figure 4.  NGC 4151 Hα light curve from Wise Observatory. Flux is in relative units. Note the small amplitude of variability despite the relatively large continuum variations shown in Fig. 3

in this object in 1989, monitored at Wise Observatory in the same way, was accompanied by Hα variations of 50% which neatly mimicked the continuum changes in the expected way (Maoz et al. 1991). The new NGC 4151 data set (including the X-rays and γ-rays which may hold more surprises) show us once again that the closer we look, the more complex AGNs appear. To reiterate my conclusion from the previous section, I think NGC 4151 is telling us that the peculiarities we see in emission line responses are the result of the ionizing continuum sometimes doing something other than what the observed continuum does.

## 3. Going To Higher Luminosities: The Steward-Wise PG Quasar Monitoring Program

Echo-mapping Seyfert galaxies has become a large industry. In the meantime, quasars have been largely neglected. This has come about due to a number of factors, mainly their faintness (and hence inaccesibility to IUE and to small telescopes), their frequent lack of narrow emission lines (still the most popular flux calibrator), and some prejudices about their (presumably long and non-paper/thesis-producing) variation and response timescales. The small amount of data that did exist on quasar emission-line variability produced controversial and sometimes contradicting interpretations (see Peterson 1993, for a review).

In 1991 we initiated a program to monitor a well-defined subsample (based on redshift, declination, apparent and absolute magnitude) of 28 of the Palomar-Green quasars (Schmidt & Green 1983). This is a fairly complete sample of UV-excess selected quasars. Note that this mode of operation differs from that which has normally been used in the field, of choosing objects with known histories of variability. Observations were carried out with the Steward Observatory 90″ and the Wise Observatory 1 m telescopes.

The main questions we wished to address were:
1. Do quasar emission lines respond to continuum changes, as seen in Seyferts?
2. What are the amplitudes and lags (and the implied BLR size) of the emission-line response?
3. Does the BLR size scale with luminosity as $R_{BLR} \propto L^{1/2}$?

The answer to the first two questions has been controversial, mainly due to the paucity of existing data. This has prevented testing the relation in the third equation, long predicted based on the overall great similarity of AGN spectra over many orders of magnitude in luminosity. Details of our program, and first results after 1.5 years of observations, appear in Maoz et al. (1994).

Figures 5 and 6 show spectra and light curves for two of the quasars in our program. The spectra in the left panels for PG 0804+762 show clearly that the Balmer lines vary in this object. Note that the spectra are plotted as they are produced by our simultaneous comparison-star long-slit method (see above) — no scaling or shifting of the spectra has been applied. (This is a good example of a quasar that cannot be flux-calibrated with the traditional [OIII] method, as it has very weak forbidden lines.)

The light curves in the right-hand panels of Figure 5 show that H$\alpha$ and H$\beta$ follow closely the optical continuum variations. The reality of the continuum and line variations in the figure is supported by the fact that the plots are composed of three independent measurements: the filled points are spectrophotometric measurements which alternate chronologically between Wise and Steward, each done relative to a *different* comparison star; the empty squares are broad-band CCD photometry as Wise, done relative to many stars in the field. Several of the quasars which underwent continuum variations also show this seemingly fast (< 1 year) line response, answering question 1 in the affirmative, and confirming previous reports of the phenomenon (e.g. Gondhalekar 1990).

On the other hand, two quasars displayed little or no line response to fairly large continuum variations. One of these, PG 1613+658, is shown in Figure 6. Judging by the hardening of this quasar's spectrum when it brightens (see Fig. 6a) the 50% flux increase we observed over a year and a half may have been accompanied by a much larger brightening in the far UV. During this period the Balmer lines showed no significant changes, as seen in the light curves (6d and 6e) and in the difference spectra (6a and 6b) where the emission lines subtract out perfectly. In the second object with constant line flux, PG 0953+414, we have compared our measurements to those taken in 1988-1990 as part of the Wise Observatory AGN monitoring program, and find that the constant line flux in the face of optical continuum changes persists over a 5-year timescale. Could this be the high-luminosity version of the NGC 4151 act we saw in the previous section (i.e. the optical continuum is not representing well the ionizing continuum)? Continued monitoring of these objects may tell.

Ignoring for the moment such complications, the apparently short emission-line lags of those quasars that did vary coherently in both lines and continuum, give us a first shot at trying to address questions 2 and 3 above. The light curves are too undersampled to get a reliable lag determination. To get a feel for what direction things are going, Kaspi (1994) has nevertheless cross-correlated the light curves and carried out Monte-Carlo simulations to determine the uncertainty range for the lag. Figure 7 (from Kaspi 1994) shows the BLR size as a function of luminosity. The three highest luminosity points are three quasars from the PG program for which a lag is obtainable. The Seyfert measurements are compiled from the literature. The $0.1 - 1\mu m$ luminosity was computed consistently for all objects, and $R_{BLR}$ is based on Balmer-line lags in all objects.

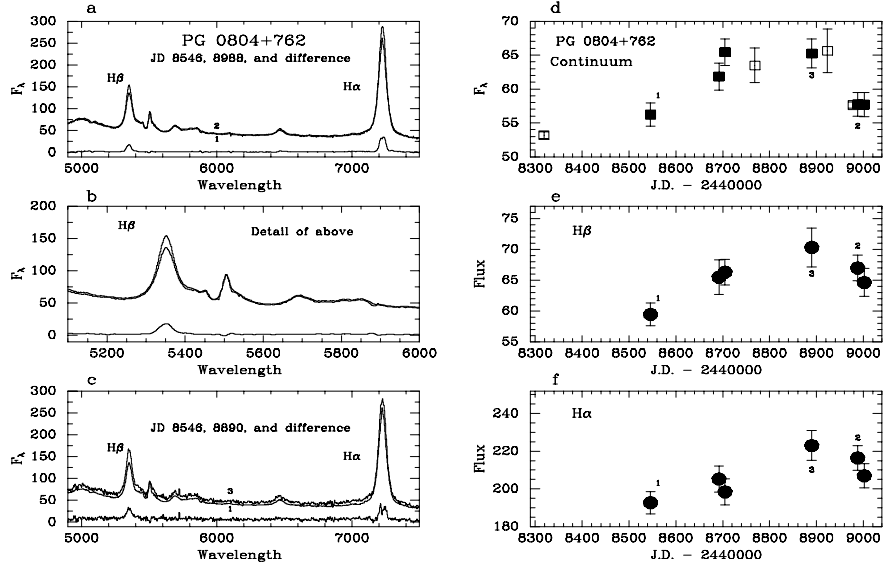

Figure 5. Spectra and light curves for PG 0804+762, from Maoz et al. (1994). Left panels show spectra at selected epochs, identified in the figure, and their difference spectra. No scaling or offsetting of the spectra has been performed. Objects with variable lines show a residual line component in the difference spectra. The panels on the right show the light curves for the $\sim 4800$Å continuum and the Balmer lines of each object. The epochs for which spectra are displayed are marked by numerals. Filled symbols are spectrophotometric measurements from Steward and Wise Observatories. Open symbols in the continuum light curves are $B$-band photometric measurements from Wise Observatory. The latter have been scaled so as to match one photometry point in every object to a spectrophotometric continuum measurement that is close in time. $F_\lambda$ is in units of $10^{-16}$ erg s$^{-1}$ cm$^{-2}$ Å$^{-1}$. Line fluxes are in units of $10^{-14}$ erg s$^{-1}$ cm$^{-2}$.

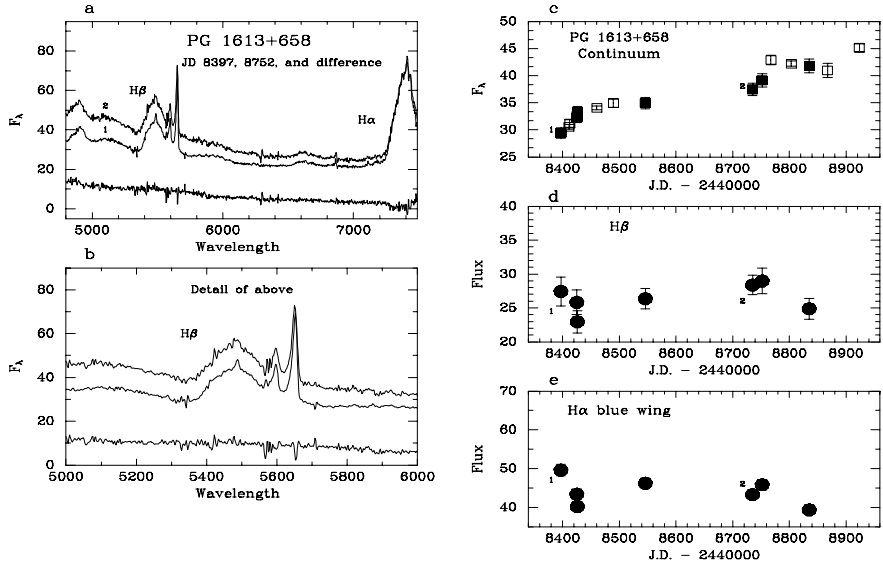

Figure 6.    Same as Fig. 5 for PG 1613+658

I again caution about the low confidence level that should be ascribed to the new quasar lags, which need to be confirmed by more data. The suggestion is, however, that the answer to question 3 above is also affirmative. If so, this will be another victory for our "big picture" of what AGNs are. On the other hand the emerging normalization of the relation in Figure 7 is problematic. Figure 7 implies

$$R_{BLR} = \left(\frac{L}{10^{41.5}\text{erg s}^{-1}}\right)^{0.5} \text{ lt} - \text{days}.$$

Assuming the line-emitting objects are sampling the gravitational potential of the central object, so that

$$M = R_{BLR} v^2 / G$$

this can be converted to a mass-luminosity relation:

$$M = 2 \times 10^7 M_\odot \left(\frac{v}{3000 \text{km s}^{-1}}\right)^2 \left(\frac{L}{10^{44}\text{erg s}^{-1}}\right)^{0.5} h^{-1},$$

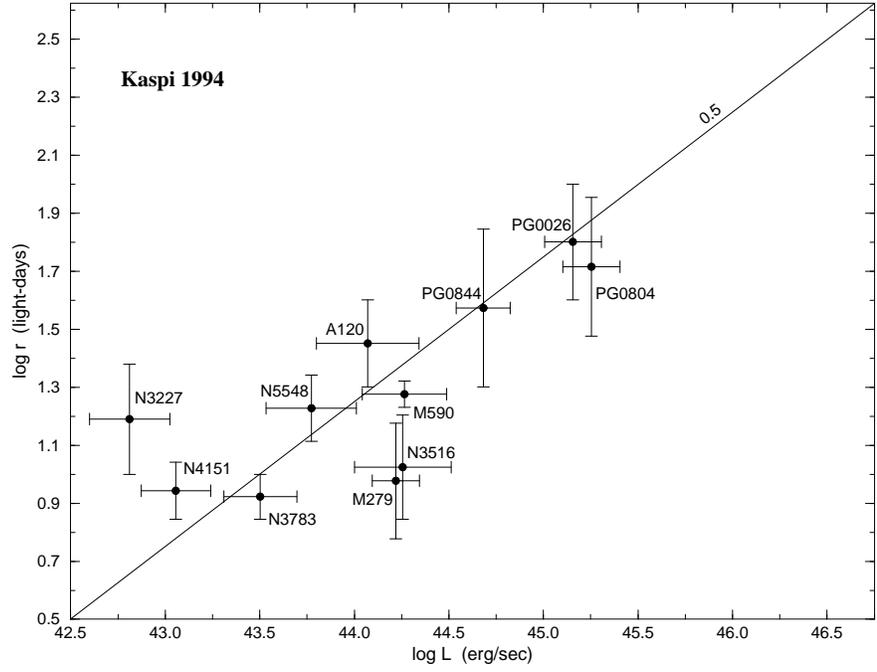

Figure 7. The BLR radius – luminosity relation, from Kaspi (1994). Luminosity between $0.1 - 1\mu$m, based on $f_\lambda(5000)$Å, and assuming $H_0 = 75$ km s$^{-1}$ Mpc$^{-1}$, $q_0 = 0.5$, and a power-law continuum spectral index $\alpha_\nu = 0.5$. All sizes based on Balmer-line lags. References: NGC 5548 – Peterson et al. (1994) and Maoz et al. (1993); Mrk 279 – Maoz et al. (1990) and Stirpe et al. (1994a); NGC 4151 – Maoz et al. (1991); Akn 120 – Peterson (1988) and Peterson and Gaskell (1991); Mrk 590 – Peterson et al. (1993); NGC 3783 – Stirpe et al. (1994b); NGC 3227 – Salamanca et al. (1994); NGC 3516 – Wanders et al. (1993); PG quasars – Maoz et al. (1994).

indicating a range of masses, going from Seyferts to high luminosity quasars, of $10^7 - 5 \times 10^8 M_\odot$. Such relatively small masses can be problematic for accretion-based energy production; The Eddington luminosity is

$$L_E = 4 \times 10^{44} \text{erg} \quad \text{s}^{-1} \left(\frac{M}{10^7 M_\odot}\right),$$

so

$$\frac{L}{L_E} = 0.1 \left(\frac{L}{10^{44} \text{erg} \quad \text{s}^{-1}}\right)^{0.5} \left(\frac{v}{3000 \text{km} \quad \text{s}^{-1}}\right)^{-2} h^{-1}.$$

Thin accretion disks can only operate at Eddington ratios of a few percent (Laor & Netzer 1989), whereas ratios of order unity or more are implied for the brightest quasars.

## 4. The Problem with Deconvolution

The ultimate purpose of echo-mapping is to obtain a "picture" of the BLR in time-delay (or time-delay / projected-velocity) space. The "picture" is the transfer function $\Psi(\tau)$ in the convolution equation relating the line and continuum light curves:

$$L(t) = \int \Psi(\tau) C(t - \tau) d\tau.$$

In the past few years, it has become common to take variability data and to attempt to invert this equation directly to recover $\Psi$ (Maoz et al. 1991; Krolik et al. 1991; Horne, Welsh & Peterson 1991; Peterson et al. 1994; Wanders & Horne 1994). Most widely used to date has been the maximum entropy inversion method. While our eagerness to finally "see" the BLR is understandable, it is essential to remember that even the best currently available variability data are very noisy. Deconvolution in the presence of noise is a risky business.

A good example of this problem is the *Hubble Space Telescope* (HST) spherical aberration (which is a completely analogous convolution problem). After its discovery there was a flurry of activity in an attempt to recover some of HST's lost performance using image reconstruction methods. It soon became apparent, however, that the reconstructions were usually dependent on the method used and the number of iterations applied. Reconstructed image features were believable only if they could clearly be seen in the raw data anyway. Worst of all, the deconvolution methods did not conserve flux, making the deconvolved data useless for any quantitative analysis. It was only for images with extremely high S/N, such as the Planetary Camera images of Jupiter, that all the different reconstructions degenerated to basically one solution. In the general case, the only quantitatively useful approach turned out to be point-spread-function modeling, e.g. subtraction of increasingly faint point sources from globular cluster images, or, for images of the central regions of galaxies, convolution of theoretical light distributions with the point-spread-function (PSF) and comparison to the observed light distributions. Such a strategy gave the range of possible images that are statistically consistent with the blurred data, and so extracted the maximum information content.

Going back to echo mapping, we are already starting off in worse shape than HST. Our light curves have low S/N, of order 10, are unevenly sampled (HST data are evenly sampled), and have a timespan comparable to the transfer functions we are looking for, forcing us to either throw away or invent much data (the HST images are much larger than the PSF). Our continuum light curves have error bars and so are imperfectly known (the HST PSF could be determined pretty well, at least empirically). Finally (see §1 and §2 above), the basic convolution assumption is not completely accurate in AGNs (in HST it is all well-understood optics).

In view of all this, I think it is unfortunate that most echo-mapping analyses done so far have taken data obtained with heroic effort and great cost, run them through a particular deconvolution algorithm, and presented us with "the" transfer function. These analyses have not addressed the truly important question which is, "Are the data inconsistent with *other* transfer functions?". The problem with every deconvolution scheme is that it gives *a* solution, not *the* solution to the convolution equation. The inverse process of modeling, on the other hand, can give the whole range of transfer functions that are consistent with the data.

The non-unique transfer functions obtained by inversion methods have been, in my opinion, overinterpreted. As an example, let us look at the H$\beta$ data for NGC 5548. Horne et al. (1991) and Peterson et al. (1994) have found that the maximum-entropy derived transfer function is peaked at 20 days, and has low-amplitude at zero lag. They interpreted this as meaning that there is little variable H$\beta$ line-emission coming from our line of sight to the nucleus, either because of the BLR geometry or due to optical-depth effects in the line.

Figure 8 shows the result of a very simple exercise. I have taken the five-year-long optical continuum light curve of NGC 5548 (Peterson et al. 1994; Korista et al. 1994), linearly interpolated it to one-day intervals, and convolved it with three different transfer functions: a delta-function peaked at 20 days, a top-hat function that is positive from 0 to 40 days, and a triangular function peaked at 0 days and decreasing to zero at 60 days. The amplitude of each of the resulting light-curves was normalized by an additive and a multiplicative constant to give a minimum $\chi^2$ when compared to the observed H$\beta$ light curve.

Figure 8 shows that the differences between the light curves produced by these very different-shaped transfer functions are minute compared to the uncertainties in the line measurements themselves. All three model light curves have similar $\chi^2$ (in fact, the top-hat and triangle give a slightly better fit to the data than the delta-function; $\chi^2 = 5.8, 5.7, 6.8$, respectively, per degree of freedom, 480 degrees of freedom). My conclusion is that these data, impressive as they are, cannot distinguish between these transfer functions, and in particular between transfer functions peaked at zero-lag and away from zero.

It can be argued that the $\chi^2$s in this example are all unacceptably high, whereas the maximimum entropy solutions have $\chi^2 = 1$. First, I note that the $\chi^2$ can be lowered by culling outliers (which are certainly in the data), revising upwards the error estimates (which are uncertain), and using a more sophisticated continuum interpolation method (I have, after all, assumed the continuum measurements are error-free). Second, the $\chi^2$ "per degree of freedom" of maximum entropy reconstructions is not really per degree of freedom, since there is a large (but difficult to quantify) number of free parameters in the maximum-entropy fit. These show up as invented variations before the monitoring period and during gaps, and "pinched" areas in the reconstructed light curves, where the model locally reaches toward the data in order to improve the fit. The example I have shown has only two free parameters (the normalization) and they were accounted for.

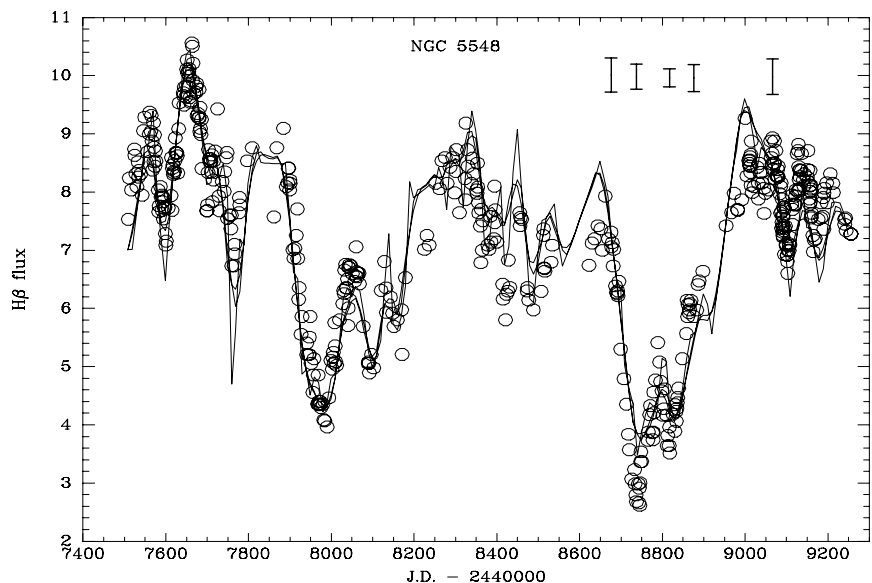

Figure 8. NGC 5548 5-year H$\beta$ light curve (empty circles) from Peterson et al. (1994 – first four years) and Korista et al. (1994 – fifth year). Some typical error bars are shown in upper right corner. The three solid lines are the convolution of the interpolated optical continuum light curves (from the same authors) with three different-shaped transfer functions: a delta-function, a top-hat, and a triangle peaked at zero lag. The resulting model light curves are difficult to distinguish, and all three fit the data equally well.

My main point is not that there is something wrong with the maximum entropy method, or that I have some better inversion method. My point is that existing data cannot distinguish between very different-shaped transfer functions. Even if a particular inversion algorithm chooses a particular transfer function solution, that does not mean that other transfer functions do not reproduce the data equally well. I showed one such example. As long as this is the case, application of inversion methods that give one transfer function as a solution to the data will produce misleading results. I was encouraged to see in this meeting that the various inversion methods that are being developed all attempt to deal with the question of assigning errors to the deconvolved transfer function. However, this still falls short of what one really wants, since the errors are generated within the a priori assumptions of a particular algorithm (e.g. smoothness and positivity in maximum entropy).

The simplest and most straightforward approach to the echo problem is to model the observed light curves, to see what is the allowable range of transfer functions that reproduce the data within measurement errors. The models need not necessarily be physical – one can use simple parametrized transfer functions, as I have in my example. Alternatively, one can include physics (e.g. photoionization) to limit some of the parameter space, and also easily include various complications in the model (such as non-linearity, see §1) which are not easily incorporated into the inversion techniques. It is only when we reach the S/N and sampling frequency of the aberrated HST images of Jupiter, that we will be in a position to safely invert echo-mapping light curves. At that point it will not matter what inversion technique we use.

**Acknowledgments.** I thank Brad Peterson for providing the NGC 5548 AGN Watch light curve prior to publication, and Shai Kaspi for providing the NGC 4151 results prior to publication, and for help with the figures.